\documentclass[
reprint,
 showkeys,
 amsmath,amsmath,amssymb,
floatfix,
]{revtex4-1}
\usepackage{graphicx}
\usepackage{bm}
\usepackage[breaklinks=true,colorlinks=true,linkcolor=blue,urlcolor=blue,citecolor=blue]{hyperref} 
\usepackage{cleveref}
\crefname{equation}{Eq.}{Eqs.}
\crefname{section}{Sec.}{ Secs.}
\crefname{figure}{Fig.}{Figs.}
\crefname{tabular}{Tab.}{Tabs.}
\creflabelformat{equation}{#2\textup{(#1)}#3}

\newcommand{\oper}{\delta\hat}
\newcommand{\toper}{\delta\tilde}
\allowdisplaybreaks
\begin{document}
\title{Atomic quadrature squeezing and quantum state transfer in a hybrid atom-optomechanical cavity with two Duffing mechanical oscillators}

\author{F. Momeni$^1$}\email{farmomeni.1392@gmail.com}
\author{M. H. Naderi$^{1,2}$}%
 
\affiliation{$^1$Department of Physics, Faculty of Science, University of Isfahan, Hezar Jerib, 81746-73441, Isfahan, Iran\\
$^2$ Quantum Optics Group, Department of Physics, Faculty of Science, University of Isfahan, Hezar Jerib, 81746-73441, Isfahan, Iran
}%

\date{\today}

\begin{abstract}
In this paper, we investigate theoretically the quantum state transfer in a laser driven hybrid optomechanical cavity with two Duffing-like anharmonic movable end mirrors containing an ensemble of identical two-level trapped atoms. The quantum state transfer from the Bogoliubov modes of the two anharmonic oscillators to the atomic mode results in the atomic quadrature squeezing beyond the standard quantum limit of 3 dB which can be controlled  by both the optomechanical and atom-field coupling strengths. Interestingly, the generated atomic squeezing can be made robust against the noise sources by means of the Duffing anharmonicity. Moreover, the results reveal that the presence of the Duffing anharmonicity provides the possibility of  transferring strongly squeezed states  between the two mechanical oscillators in a short operating time and with a high fidelity.

\end{abstract}

\maketitle


\section{\label{sec:int}Introduction}
Over the past decade,  cavity optomechanical systems have received a great deal of research attention from both theoretical and experimental points of view because they not only provide the opportunity to gain new insights into the quantum-to-classical transition~\cite{T1}, but also have been demonstrated to possess remarkable applications in various fields including  the ground-state cooling of mechanical oscillators~\cite{T2,T3,T4,T5,T6}, generation of the mechanical and optical nonclassical states~\cite{T7,T8,T9}, and coherent state transfer between the cavity and mechanical modes~\cite{T10,T11}, to mention a few.  In a prototypical cavity optomechanical system, the mechanical motion of a movable mirror interacts with the radiation field of a high-finesse cavity via the radiation pressure force which, to the lowest order, is linearly proportional to the displacement of the movable mirror~\cite{T12}. By realizing many-photon strong couplings between the mechanical resonator and optical cavity~\cite{T13,T14} and assuming that the rotational wave approximation (RWA) is valid, two types of interactions can be distinguished~\cite{T12}: the parametric down-conversion interaction which is behind all forms of bipartite Gaussian entanglement~\cite{T15}, and the beam-splitter like interaction which is responsible for quantum state transfer between the cavity field and the mechanical oscillator~\cite{T16,T17a,T17}.

 Photons are the fastest and most robust carriers of quantum information~\cite{T18}, however, their localization and storage are very challenging. For this reason, much attention has been paid to the quantum state transfer between the intra-cavity field and movable mirror as a quantum memory element.
So far, many works have been devoted to the problem of state transferring  between optical and microwave cavities with the help of an intermediary mechanical oscillator~\cite{T10,T19,T20,T21,T22,T23}. In addition, it has been shown~\cite{T24} that the cavity field can also be a mediator for transferring the quantum state between two distant mechanical oscillators.  In \cite{T16} it has been shown that the mechanical squeezed state can be obtained by transferring the quantum correlation between two input optical fields to a single macroscopic oscillator. Also, it was pointed out that after eliminating adiabatically the linearized cavity field, the transfer of the single optical mode correlation into a pair of  macroscopic oscillators can pull them into a two-mode squeezed state. Furthermore,  the quantum state transfer between two distant mechanical oscillators in two coupled  optomechanical cavities before and after the arrival of the mediating cavity modes to  the steady state has been investigated,  respectively, in~\cite{T17a} and~\cite{T17}. 

A common feature of the above-cited investigations is that they treat approximately the quantum mechanical oscillator as a harmonic one, as its intrinsic anharmonicity (nonlinearity) is considered small enough to be ignored. In recent years, the realization as well as the enhancement of the anharmonic (nonlinear) regime have been experimentally demonstrated in some settings, including mechanical resonators based on graphene and carbon nanotubes~\cite{T25,T26}, levitated nanoparticles~\cite{T27}, and optoelectromechanical systems~\cite{T28}. On the other hand, it has been found that anharmonicity gives rise to some interesting quantum phenomena.  For example, the investigation of the quantum and classical dynamics of an anharmonic (nonlinear) oscillator in phase space shows that a decoherence-induced state reduction results in a quantum-to-classical transition~\cite{T29}. In~\cite{T30} a theoretical scheme has been proposed to generate periodic and chaotic optical signals in an electro-optomechanical system via the Duffing-type of mechanical anharmonicity. Besides, quantum anharmonic oscillators provide new possibilities for quantum state generation and manipulation in mechanical systems. In this direction, it has been shown~\cite{T31,T32,T33} that anharmonicity is a resource to create nonclassical quantum states such as entangled and squeezed states of a mechanical oscillator. In the context of optomechanical self-oscillations, a theoretical proposal has been made~\cite{T34} to achieve the steady-state sub-Poissonian phonon statistics in an optomechanical cavity, by using the intrinsic anharmonicity of the mechanical oscillator. Furthermore, it has been recently proposed a protocol to estimate the anharmonicity of a quantum mechanical oscillator in an optomechanical cavity~\cite{T35} 
in order to explore its contribution to the dynamics and its impact on the experimental results.

In this work, we are going to investigate theoretically the role of the intrinsic mechanical anharmonicity on quantum state transfer in a hybrid atom-optomechanical cavity. For this purpose, we consider an optomechanical cavity with two moving end mirrors which are modeled as two stiffening Duffing-like anharmonic quantum oscillators. The cavity which is driven by a strong input laser contains an ensemble of identical two-level atoms. 
 Since the intrinsic (geometrical) nonlinearity of a sub-gigahertz micro or nano-mechanical resonator is usually very weak in the regime of very small oscillation amplitudes \cite{T36}, we follow Ref.~\cite{T33} to produce a strong mechanical  nonlinearity. We investigate the state transfer between the mechanical modes as well as between the collective mechanical Bogoliubov modes and the atomic mode before the cavity field arrives at the steady state. To describe the quantum state transfer between any two modes of the system, we adopt the so-called "hybrid" state transfer scheme~\cite{T22}. We show that the quantum state transfer between the mechanical oscillators depends on the atom-field coupling and, therefore, this quantity can be used to control the distribution of the quantum information. It is shown that the presence of the Duffing anharmonicity causes the RWA to be valid in a wide range of the optomechanical coupling rates. This effect makes it possible to increase the optomechanical coupling strength significantly, and thus transmit strongly squeezed states in a short operating time and with a high fidelity. Furthermore, we show that the state transfer from the mechanical Bogoliubov mode to the atomic mode may lead to the atomic quadrature squeezing  beyond the standard quantum limit of 3 dB. This squeezing, which is produced before the cavity field reaches the steady state, is controlled by  the optomechanical and atom-field coupling strengths and can show good robustness against the thermal bath's temperature and the cavity-field damping.

The paper is structured as follows. In \cref{sec:2nd}, we describe the model system under consideration,
give the linearized quantum Langevin equations for quantum fluctuations, and the dynamics of the symmetrized covariance matrix of  quadrature fluctuation operators. In \cref{secIII}, we study the squeezing transfer from the collective mechanical mode to the atomic mode, and quantum state transfer between the mechanical oscillators. Finally, conclusions are summarized in \cref{conc}. In addition, some mathematical details are presented in the Appendix.

\section{\label{sec:2nd} Theoretical discription of the system}
\begin{figure}[h]
\includegraphics[scale=0.67]{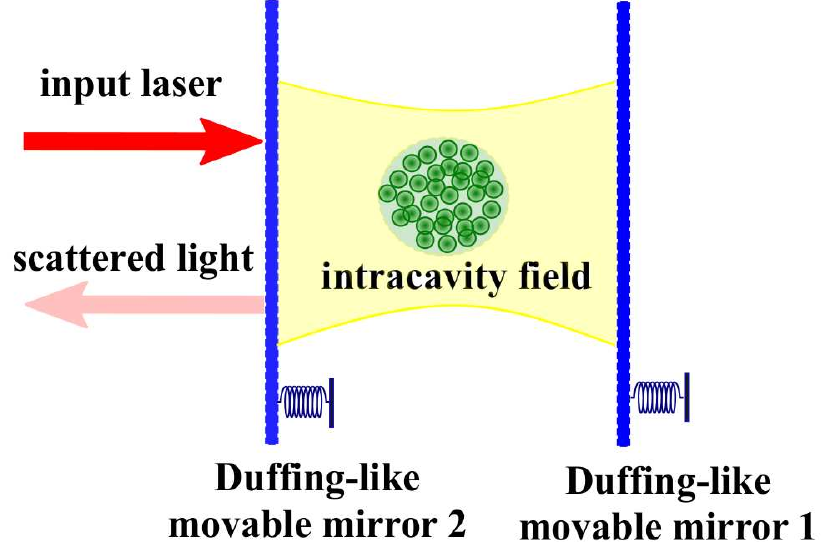}
\caption{\label{fig1}Schematic illustration of the hybrid optomechanical system discussed in the text. Two Duffing-like vibrating mirrors are commonly coupled to the single-mode radiation field of a strongly driven cavity which contains an ensemble of identical two-level atoms.}
\end{figure}

As shown in \cref{fig1}, we consider an optomechanical cavity with two movable mirrors commonly coupled to a single-mode intracavity field with resonance frequency $\omega_c$ and cavity decay rate $\kappa$. Each of the two mirrors is modeled as a single-mode Duffing-like anharmonic quantum mechanical oscillator with resonance frequency $\omega_j$, effective mass $m_j$, energy decay rate $\gamma_j$, and Duffing anharmonicity parameter $\lambda_j$ ($j=1,2$). The single-mode cavity field is coherently driven by a strong laser field with input power $P_{\text{in}}$, frequency $\omega_L$, and amplitude $\vert \varepsilon\vert=\sqrt{\frac{2{\kappa} {P_{\text{in}}}}{\hbar {\omega_L}}}$ through the partially transmitting left mirror. The cavity also contains an ensemble of $N$ identical two-level atoms, each of which is described by a ground state $\vert g_i\rangle$, an excited state $\vert e_i\rangle$, transition frequency $\omega_e$, and decay rate $\gamma_e$. 

The total Hamiltonian of the system  in a frame rotating at the laser frequency $\omega_L$ can be written as ($\hbar \equiv 1$)
\begin{align}\label{fH}
\hat{H}&= \hat{H}_{op}+\hat{H}_{at},
\end{align}
with
\begin{align}\label{Hop}
\hat{H}_{op}&= (\omega_c-\omega_L)\hat{a}^\dagger\hat{a}+i(\varepsilon \hat{a}^\dagger-\varepsilon^*\hat{a})\nonumber\\&+\sum_{j=1,2}\{\omega_j\hat{b}^\dagger_j\hat{b}_j+\frac{\lambda_j}{2}(\hat{b}_j+\hat{b}^\dagger_j)^4+(-1)^j g_j\hat{a}^\dagger\hat{a}(\hat{ b}_j+\hat{b} ^\dagger_j)\},\\
\hat{H}_{at}&= \frac{\Delta_e}{2} \sum_{i=1}^N\hat{\sigma}_i^{z}+\frac{\bar{\eta}}{2}(\hat{a}\sum_{i=1}^N\hat{\sigma}_i^{+}+\hat{a}^\dagger\sum_{i=1}^N\hat{\sigma}_i^{-}),
\end{align}
where $\hat{a},\hat{b}_2$, and $\hat{b}_1$ ($\hat{a}^\dagger,\hat{b}_2^\dagger,\hat{b}_1^\dagger$) denote, respectively,  the annihilation (creation) operators of the cavity mode, the mechanical motion of the left moving mirror, and that of the right one. The operators $\hat{\sigma}_i^{z}$ and $\hat{\sigma}_i^{\pm}$, which are used to describe the $i$th two-level atom in the ensemble, are the spin-1/2 Pauli matrices defined by $\hat{\sigma}_i^{z}=1/2(\vert e_{i}\rangle \langle e_{i}\vert-\vert g_{i}\rangle \langle g_{i}\vert)$ and $\hat{\sigma}_i^{+}=\vert e_{i}\rangle \langle g_{i}\vert=(\hat{\sigma}_i^{-})^\dagger$. In the Hamiltonian $\hat{H}_{op}$,  the first term accounts for the cavity-mode energy and the second term describes the coupling between the cavity mode and the driving laser. In addition, the first and second terms in the brackets correspond to the nonlinear Hamiltonian of the $j$th Duffing-like mechanical resonator. Finally, the third term in the brackets denotes the interaction between the optical field and the $j$th mechanical oscillator via the radiation pressure force with single-photon coupling strength $g_j=\frac{{\omega}_{c}}{L}\sqrt{\frac{\hbar}{2m_j \omega_j}}$ ($L$ being the length of cavity). In Hamiltonian $\hat{H}_{at}$, the first  term represents the energy of the atomic ensemble with ${\Delta_e}={\omega_e}-\omega_L$,  and  the second term describes the interaction between the atomic ensemble and the intracavity field  with the averaged coupling strength $\bar{\eta}=\sum_{i=1}^{N}\eta_i/N$ in which $\eta_i$ is the coupling strength of the $i$th two-level atom with the cavity mode.

In the low-excitation limit, i.e., when the number of atoms in the excited state is much lower than the total number of atoms, the atomic ensemble can be treated as a bosonic mode  with annihilation and creation operators $\hat{c}$ and  $\hat{c}^\dagger$ which are related to the Pauli matrices via the Holstein-Primakoff representation \cite{T38}:
\begin{subequations}
\begin{align}
&\frac{1}{\sqrt{N}}\sum_{i=1}^{N} \hat{\sigma}_i^{-}=\hat{c},\\
&\frac{1}{\sqrt{N}}\sum_{i=1}^{N} \hat{\sigma}_i^{+}=\hat{c}^\dagger,\\
&\sum_{i=1}^{N} \hat{\sigma}_i^{z}=2\hat{c}^\dagger\hat{c}-N.
\end{align}
\end{subequations}
Therefore, the Hamiltonian $\hat{H}_{at}$ can be rewritten as
\begin{align}
\hat{H}_{at}=\Delta_e \hat{c}^\dagger\hat{c}+{\eta}_e(\hat{a}\hat{c}^\dagger+\hat{a}^\dagger\hat{c}),
\end{align}
where $\eta_e=\sqrt{N}\bar{\eta}/2$ is the enhanced atom-field coupling strength.

Taking into account the   fluctuation-dissipation processes affecting the cavity-field, atomic,  and mechanical modes, the dynamics of the hybridized optomechanical system is determined by the following set of nonlinear quantum Langevin equations 
\begin{subequations}
\begin{align}
\dot{\hat{a}}=&-i(\omega_c-\omega_L)\hat{a}+ig_1\hat{a}(\hat{b}_1+\hat{b}^\dagger_1)-ig_2\hat{a}(\hat{b}_2+\hat{b}^\dagger_2)\nonumber\\&-i\eta_e\hat{c}+\varepsilon-\kappa\hat{a}+\sqrt{2\kappa}\hat{a}_{in},\label{6a}\\
\dot{\hat{c}}=&-i\Delta_e\hat{c}-i\eta_e\hat{a}-\gamma_e\hat{c}+\sqrt{2\gamma_e}\hat{c}_{in},\label{6b}\\
\dot{\hat{b}}_j=&-i\,\omega_j \hat{b}_j-2i\,\lambda_j(\hat{b}_j+\hat{b}^\dagger_j)^3-i(-1)^j\,g_j\hat{a}^\dagger\hat{a}\nonumber\\&-\gamma_j \hat{b}_j+\sqrt{2\gamma_j}\hat{b}_{in,j},\label{6c}
\end{align}
\end{subequations}
 where the optical input vacuuum noise  $\hat{a}_{in}$ and  the atomic ensemble vacuum input noise $\hat{c}_{in}$, with zero mean values, are characterized by the nonvanishing Markovian correlation functions  $\langle \hat{a}_{in}(t)\hat{a}_{in}^\dagger(t')\rangle=\delta(t- t')$ and $\langle \hat{c}_{in}(t)\hat{c}_{in}^\dagger(t')\rangle=\delta(t- t')$, respectively \cite{T39}. Also, in the limit of high mechanical quality factor, i.e., $Q_j=\omega_j/\gamma_j\gg 1$,  the Brownian noise operator of the mechanical oscillator $\hat{b}_{j,in}$ satisfies the nonvanishing Markovian correlation functions $\langle \hat{b}_{in,j}(t)\,\hat{b}_{in,j}^\dagger(t')\rangle =(1+\bar{n}_{th,j})\delta(t-t')$, $\langle\hat{b}_{in,j}^\dagger(t)\hat{b}_{in,j}(t')\rangle=\bar{n}_{th,j} \,\,\delta(t-t')$ \cite{T40} where $\bar{n}_{th,j}=[\text{exp}(\hbar \omega_j/k_B T_j)-1]^{-1}$ is the mean number of thermal phonons of the $j$th mechanical oscillator at temperature $T_j$ with $k_B$  being the Boltzmann constant.
 
\begin{figure}[h]\centering
\includegraphics[scale=0.42]{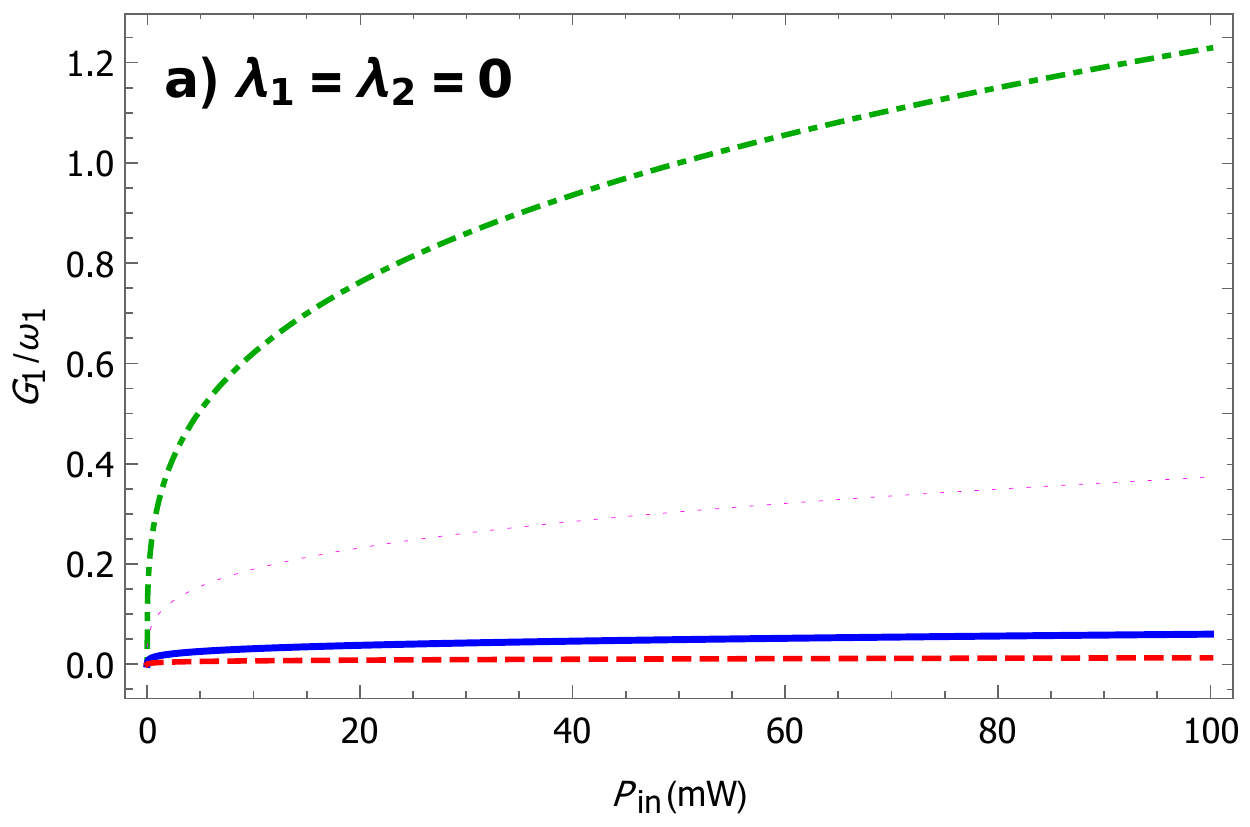}\par
\includegraphics[scale=0.42]{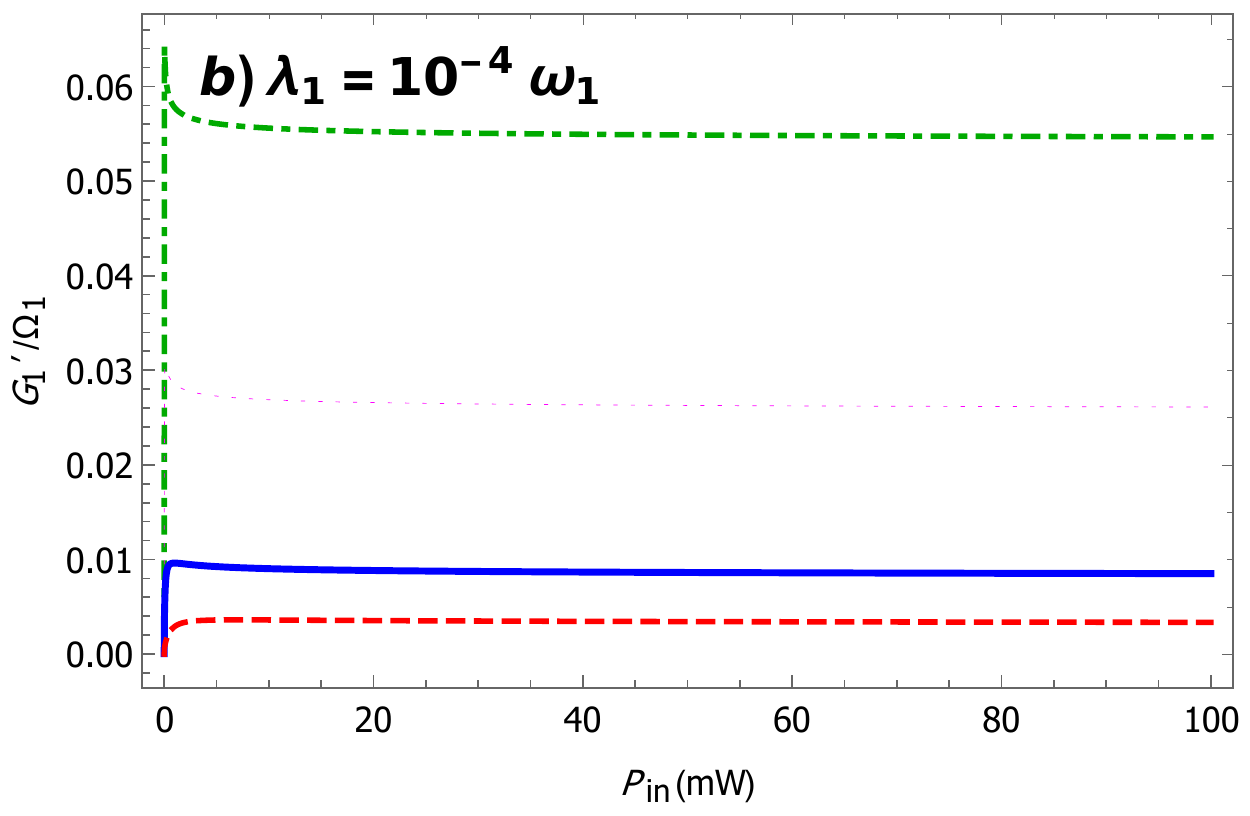}
\caption{\label{fig2}(a) The normalized effective optomechanical coupling rate $G_1/\omega_1$ in the absence of the Duffing anharmonicity, and (b) the normalized transformed optomechanical coupling rate $G'_1/\Omega_1$ in the presence of the Duffing anharmonicity versus the input laser power $P_{\text{in}}$ for different values of the single-photon optomechanical coupling strength $g_1$: $g_1\approx1.92\times 10^{-6}\omega_1$ (red dashed line),  $g_1\approx1.36\times 10^{-5}\omega_1$ (blue solid line), $g_1\approx1.36\times 10^{-4}\omega_1$ (magenta dotted line),  $g_1\approx6.07\times 10^{-4}\omega_1$ (green dot-dashed  line). Here, we have used the following set of experimentally realizable parameters \cite{T41,T42,T43,T44}: driving laser wavelength $\lambda_L=810\text{nm}$, cavity damping rate  $\kappa = \pi\times 10^5  s^{-1}$ ($\kappa=0.01\omega_1$), mechanical resonance frequency $\omega_1=\omega_2=2\pi\times5\text{MHz}$, mechanical damping rate $\gamma_1=\gamma_2=10^{-5}\omega_1$, Duffing parameter $\lambda_2=10^{-4}\omega_2$, and $g_2=g_1$. In addition, we have set  $\Delta=\Delta_e=\Omega_2=\Omega_1$ and have used the following parameters for the atomic ensemble:  enhanced atom-field coupling $\eta_e=0.3\omega_1$ and atomic decay rate $\gamma_e=\gamma_1$.
  }
\end{figure}

 In the regime of strong external driving field one can exploit the mean-field approximation in which the quantum operators are expressed as the sum of their steady-state mean values and small quantum fluctuations, i.e., $\hat{o}=\langle \hat{o}\rangle_s+\oper o$ with $\langle \oper{o}^\dagger \oper{o}\rangle_s/\langle \hat{o}^\dagger\hat{o}\rangle_s\ll 1$ ($o=a, c, b_1, b_2$). In this manner, Eqs.~(\ref{6a})-(\ref{6c}) lead to a set of linearized quantum Langevin equations for the quantum fluctuation operators, $\oper o$, from which one can deduce the following linearized Hamiltonian 
\begin{align}\label{Hlin}
\hat{H}_L&= \Delta\oper a^\dagger\oper a+\Delta_e \oper{c}^\dagger\oper{c}+{\eta}_e(\oper{a}\oper{c}^\dagger+\oper{a}^\dagger\oper{c})\nonumber\\&\hspace*{0.2cm}+\sum_{j=1,2}\{(\omega_j+2\Lambda_j)\oper b^\dagger_j\oper b_j+\Lambda_j(\oper b_j^2+\oper b^{\dagger 2}_j)\nonumber\\&\hspace*{2cm}+(-1)^j G_j(\oper a+\oper a^\dagger)(\oper b_j+\oper b^\dagger_j)\},
\end{align}
where $\Delta=\omega_c-\omega_L-2g_1 \,\beta_{1,s}+2g_2\,\beta_{2,s}$, $G_j=g_j\alpha_s$, and $\Lambda_j=3\lambda_j(1+4\beta_{j,s}^2)$  are the effective detuning of the cavity field, the enhanced optomechanical coupling rate, and  the enhanced Duffing parameter of the movable mirror $j$, respectively. Also, the steady-state amplitude of the intracavity field $\alpha_s=\langle \hat{a}\rangle_s$ and the steady-state oscillation amplitude $\beta_{j,s}=\text{Re}(\langle\hat{b}_j\rangle_s)$  obey the following set of nonlinear algebraic equations
\begin{subequations}
\begin{align}
&\alpha_s=\frac{|\varepsilon|}{\sqrt{(\,\Delta-\frac{\eta_e^2\Delta_e}{\Delta_e^2+\gamma_e^2})^2+(\kappa+\frac{\eta_e^2\gamma_e}{\Delta_e^2+\gamma_e^2}})^2},\\
&16\frac{\lambda_j}{\omega_j}\beta_{j,s}^3+(1+12\frac{\lambda_j}{\omega_j}+\frac{\gamma_j^2}{\omega_j^2})\beta_{j,s}+(-1)^j\frac{g_j}{\omega_j}\alpha_s^2=0.
\end{align}\end{subequations}
To get a beam-splitter type of interaction, which is responsible for quantum state transfer between  different modes, it is convenient to diagonalize the quadratic mechanical part of $\hat{H}_L$ by introducing the mechanical Bogoliubov mode $\toper B_j=\mu_j\oper b_j+\nu_j\oper b^\dagger_j$ in which
\begin{align}
&\mu_j=\cosh(r_j), \quad \nu_j=\sinh(r_j),
\end{align}
with $r_j=\frac{1}{4}\log(1+\frac{4\Lambda_j}{\omega_j})$.  Under this transformation, the Hamiltonian of Eq.~(\ref{Hlin}) is transformed into
\begin{align}\label{Hfin}
\hat{H}_L&= \Delta\oper{a}^\dagger\oper{a}+\Delta_e \oper{c}^\dagger\oper{c}+\Omega_1\toper B^\dagger_1\toper B_1+\Omega_2\toper B^\dagger_2\toper B_2\nonumber\\&+{\eta}_e(\oper{a}\oper{c}^\dagger+\oper{a}^\dagger\oper{c})-G_1'(\oper{a}^\dagger +\oper{a})(\toper B_1+\toper B^\dagger_1)\nonumber\\&+G_2'(\oper{a}^\dagger +\oper{a})(\toper B_2+\toper B^\dagger_2),
\end{align}
in which $\Omega_j=\omega_j\,e^{2r_j}$  and $G'_j=G_j\,e^{-r_j}$. Performing the RWA by dropping rapidly oscillating terms, which is justified when $G_1',G_2'\ll \Omega_1,\Omega_2,\Delta$, \cref{Hfin} can be approximated by a ``beam-splitter like'' Hamiltonian which is responsible for the state transferring between the mechanical Bogoliubov modes through the optical mode. But beyond the RWA, due to the unavoidable contribution of the counter rotating terms, quantum state transfer cannot be done perfectly. It is important to note that the required condition for the validity of the RWA depends on the value of $r_j$; in the absence of the Duffing anharmonicity ($r_j=0$)  the condition is $G_j /\omega_j\ll 1$, while non-zero  Duffing parameter ($r_j>0$) extends the range of the values of the enhanced optomechanical coupling strength $G_j$ over which one can perform state transfer with minimal impact of the counter rotating terms $\oper a\,\oper{ B}_j $ and $\oper a ^\dagger\oper{ B}_j^\dagger$ ($j=1,2$). In \cref{fig2}, we have plotted the normalized effective optomechanical coupling rate $G_1/\omega_1$(\cref{fig2}(a)) and the normalized transformed optomechanical coupling rate $G'_1/\Omega_1$(\cref{fig2}(b)) versus the input laser power $P_{\text{in}}$ for different values of the single-photon optomechanical coupling rate $g_1$. From the comparison of these two figures, it can be concluded that in the absence of the Duffing anharmonicity with increasing $P_{\text{in}}$ and also increasing $g_1$, the RWA loses its validity. This is while the existence of the Duffing anharmonicity causes the RWA to remain valid over a wide range of  values of $P_{\text{in}}$ and $g_1$.

By tuning $\Delta=\Delta_e=\Omega_2=\Omega_1$, within the RWA  and in the interaction picture with respect to the free Hamiltonian $\hat{H}_0=\Omega_1(\oper{a}^\dagger\oper{a}+\oper{c}^\dagger\oper{c}+\toper B^\dagger_1\toper B_1+\toper B^\dagger_2\toper B_2)$, the linearized quantum Langevin equations for the fluctuation operators read as
\begin{subequations}
\begin{align}
\delta\dot{\tilde{a}}=&iG_1'\toper B_1-iG_2'\toper B_2-i\eta_e\toper{c}-\kappa \toper a+\sqrt{2\kappa}\tilde{ a}_{in},\label{eq1}\\
\delta\dot{\tilde{c}}=&-i\eta_e\toper{a}-\gamma \toper c+\sqrt{2\gamma}\tilde{ c}_{in},\label{eq11}\\
\delta\dot{\tilde{B}}_j=&-i(-1)^j\,G'_j\toper{a}-\gamma \toper B_j+\sqrt{2\gamma}\tilde{ B}_{in,j},\label{eq111}
\end{align}
\end{subequations}
where ${\tilde{ o}=e^{i\Omega_1\,t}\hat{o}}$ (${o=\delta a, a_{in},\delta c, c_{in},\delta B_j, B_{in,j}}$) and, for the sake of simplicity, we have assumed that $\gamma_1=\gamma_2=\gamma_e=\gamma$. Furthermore, $\hat{B}_{in,j}=\mu_j \hat{b}_{in,j}+\nu_j \hat{b}_{in,j}^\dagger$ is the input Brownian noise of the Bogoliubov mode of the  $j$th mechanical oscillator with the following correlation functions:
\begin{subequations}
\begin{align}
&\langle \tilde{ B}_{in,j}(t)\,\tilde{ B}_{in,j}^\dagger(t')\rangle =(1+\bar{n}_{r,j})\delta(t-t'),\label{tcf1}\\
&\langle \tilde{ B}_{in,j}^\dagger(t)\,\tilde{ B}_{in,j}(t')\rangle=\bar{n}_{r,j}\delta(t-t'),\label{tcf2}\\
&\langle \tilde{ B}_{in,j}(t)\,\tilde{ B}_{in,j}(t')\rangle =e^{2i\Omega_1\,t}\mu_j\,\nu_j(1+2\bar{n}_{th,j})\delta(t-t')\nonumber\\
&\hspace*{2.7cm}=\langle \tilde{ B}_{in,j}(t)\,\tilde{ B}_{in,j}(t')\rangle^*,\label{tcf3}
\end{align}\end{subequations}
with $\bar{n}_{r,j}=(\mu_j^2+\nu_j^2)\bar{n}_{th,j}+\nu_j^2$. We also define the vector of quadrature fluctuation operators as\\ $\oper{\vec{ U}}=(\oper x_c,\oper y_c,\oper x_a,\oper y_a,\oper Q_1,\oper P_1,\oper Q_2,\oper P_2)^T$ in which   ${\oper x_c=\frac{\toper c+\toper c^\dagger}{\sqrt{2}}}$ and ${\oper y_c=\frac{\toper c-\toper c^\dagger}{\sqrt{2}i}}$ refer to the atomic ensemble quadratures fluctuations, ${\oper x_a=\frac{\toper a+\toper a^\dagger}{\sqrt{2}}}$ and ${\oper y_a=\frac{\toper a-\toper a^\dagger}{\sqrt{2}i}}$ represent the cavity-field quadratures fluctuations, and  ${\oper Q_j=\frac{\toper B_j+\toper B_j^\dagger}{\sqrt{2}}}$ and ${\oper P_j=\frac{\toper B_j-\toper B_j^\dagger}{\sqrt{2}i}}$ ($j=1,2$) denote the mechanical quadratures fluctuations. Then, by using of Eqs.~(\ref{eq1})-(\ref{eq111}), it can be shown (see the Appendix) that $\oper{\vec{U}}$ obeys the following relation
\begin{equation}\label{key1}
\oper{\vec{ U}}(t)=\mathbf{M}(t).\oper{\vec{ U}}(t_0=0)+\hat{\vec{\Gamma}}(t),
\end{equation}
in which $\mathbf{M}$ is a  $8\times 8$  matrix and $\hat{\vec{\Gamma}}$ denotes the vector of noises whose explicit forms are given in the Appendix.  Also, since  the noises are assumed to be Gaussian, the first moment of the quadratures of each mode is made by extracting the corresponding elements from \begin{small}$\langle \oper{\vec{ U}}(t)\rangle=\mathbf{M}(t).\langle\oper{\vec{ U}}(t_0=0)\rangle$\end{small}. 

 Furthermore, we define the symmetrized covariance matrix for the entire system as $\mathbf{V}=\frac{1}{2}\langle \oper{\vec{ U}}\oper{\vec{ U}}^T+\oper{\vec{ U}}^T\oper{\vec{ U}}\rangle-\langle\oper{\vec{ U}}\rangle\langle\oper{\vec{ U}}^T\rangle$ which, by means of \cref{key1}, is obtained as
\begin{align}
&\mathbf{V}(t)=\mathbf{M}(t).\mathbf{V}(0).\mathbf{M}^T(t)+\int_0^td\tau\mathbf{M}(\tau).\mathbf{N}.\mathbf{M}^T(\tau),\label{k4}
\end{align}
in which $\mathbf{V}(0)$ is the symmetrized covariance matrix at the initial time $t=0$. Also, we have defined the \textit{diffusion} matrix $\mathbf{N}$ as the symmetrized noise covariance matrix, i.e., $\mathbf{N}=\frac{1}{2}\langle \hat{\vec{n}}\hat{\vec{n}}^T+\hat{\vec{n}}^T\hat{\vec{n}}\rangle$ with $\hat{\vec{n}}$ being the vector of noises defined in the Appendix. Employing the correlation functions of Eqs.~(\ref{tcf1})-(\ref{tcf2}),  $\mathbf{N}$  can be written as $\mathbf{N}_0+\mathbf{N}_t$ where $\mathbf{N}_0$ and $\mathbf{N}_t$ are, respectively, time-independent and time-dependent parts of $\mathbf{N}$ whose explicit forms are presented in the Appendix. For the next calculation, with the condition $\gamma\mu_j\nu_j(1+2\bar{n}_{th,j}),\,G_j^{'}\ll\Omega_1$, we will neglect the time-dependent part $\mathbf{N}_t$  under RWA. 

\section{state transfer in the system\label{secIII}}
In this section, we are going to investigate the quantum state transfer between the collective mechanical mode and the atomic mode (subsection \ref{sec31}) as well as between the mechanical oscillators (subsection \ref{sec32})  before the cavity field reaches its steady state. In addition, we consider the so-called ''hybrid'' scheme \cite{T22} for describing state transfer wherein the corresponding coupling rates are assumed to be turned on simultaneously.

In order to make the transfer process clear, we temporarily ignore the dissipation factors in the dynamics, i.e., $\kappa,\gamma=0$. By monitoring the atomic, cavity-field, and mechanical modes over time, it can be concluded that state transfer between the mechanical oscillators can be achieved immediately after the operating time $t_s^{(0)}=\pi/\sqrt{G'^{2}_1+G'^{2}_2+\eta_e^2}$.  By applying \cref{key1} and the definitions of the quadrature fluctuation operators, the fluctuation operators at $t_s^{(0)}$ read as
\begin{small}\begin{subequations}
\begin{align}
&\toper a(t_s^{(0)})=-\oper a(0),\label{a1}\\
&\toper c(t_s^{(0)})=\frac{1}{G'^{2}_1+G'^{2}_2+\eta_e^2}\Big(2G_1'\eta_e\oper{B}_1(0)-2G_2'\eta_e\oper{B}_2(0)\nonumber\\ &\hspace*{3cm}+(G'^{2}_2+G'^{2}_1-\eta_e^2)\oper{c}(0)\Big),\label{c1}\\
&\toper B_1(t_s^{(0)})=\frac{1}{G'^{2}_1+G'^{2}_2+\eta_e^2}\Big(2G_1'G_2'\oper{B}_2(0)+G'_1\eta_e\oper{c}(0)\nonumber\\ &\hspace*{3cm}+(G'^{2}_2-G'^{2}_1+\eta_e^2)\oper{B}_1(0)\Big),\label{B1}\\
&\toper B_2(t_s^{(0)})=\frac{1}{G'^{2}_1+G'^{2}_2+\eta_e^2}\Big(2G_1'G_2'\oper{B}_1(0)-G'_2\eta_e\oper{c}(0)\nonumber\\ &\hspace*{3cm}+(G'^{2}_1-G'^{2}_2+\eta_e^2)\oper{B}_2(0)\Big).\label{B2}
\end{align}
\end{subequations}\end{small}From the above equations, it is clear  that although the cavity field retains its initial state in which it was prepared, the final states of the mechanical and atomic modes depend on the coupling parameters $G_1',G_2',\eta_e$. As can be seen, the state swapping between the mechanical Bogoliubov modes can be possible  when the enhanced atom-field coupling strength $\eta_e$ is much weaker than the effective optomechanical coupling strength $\sqrt{G'^{2}_1+G'^{2}_2}$. In this case, Eqs.~(\ref{c1})-(\ref{B2}) are reduced to 
\begin{small}\begin{subequations}\begin{align}
&\toper c(t_s^{(0)})=\oper c(0),\\
&\toper B_1(t_s^{(0)})=\frac{\Big(2G_1'G_2'\oper{B}_2(0)+(G'^{2}_2-G'^{2}_1)\oper{B}_1(0)\Big)}{G'^{2}_1+G'^{2}_2},\\
&\toper B_2(t_s^{(0)})=\frac{\Big(2G_1'G_2'\oper{B}_1(0)+(G'^{2}_1-G'^{2}_2)\oper{B}_2(0)\Big)}{G'^{2}_1+G'^{2}_2},
\end{align}\end{subequations}\end{small}It is clearly seen that the atomic ensemble mode has regained its initial state, and  the mechanical modes will completely exchange their states when they are coupled to the cavity mode with the same strength ($G^{'}_2=G_1'$). In fact, the atom-field coupling can be considered as a control parameter for the quantum state transfer between the two mechanical oscillators. When $\eta_e\rightarrow 0$, the perfect state transfer  between the two mechanical oscillators can occur while for $\eta_e=\sqrt{G_1'^2+G_2'^2}$ the transfer of state between the atomic ensemble and the collective mechanical mode ${\frac{G_1'\toper{B}_1-G_2'\toper{B}_2}{\sqrt{G'^{2}_1+G'^{2}_2}}}$ becomes possible. On the other hand, when the atomic ensemble is strongly coupled to the intracavity field (${\eta_e\gg G_1',G_2'}$) after the optimum time $t_s^{(0)}$ each mode retains its initial state so that the state transfer is totally inhibited. It should be pointed out that it can easily be shown that for ${\eta_e\gg G_1',G_2'}$ the transfer of the quantum state  between mechanical oscillators can also be possible provided that $\Delta=\Delta_e=\Omega_1\pm \eta_e$ is established.

In general, the presence of the noise sources seriously disrupts the state transfer  between different modes, unless in the many-photon strong coupling condition $\gamma\bar{n}_{r,j}, |\kappa-\gamma|/2\ll \sqrt{G'^{2}_1+G'^{2}_2}$. Also, in the presence of dissipation processes, the swapping time is modified to $t_s=\pi/\mathcal{G}$ with $\mathcal{G}=\sqrt{G'^{2}_1+G'^{2}_2+\eta_e^2-(\kappa-\gamma)^2/4}$.

\subsection{atomic quadrature squeezing\label{sec31}}
In this subsection, we examine and compare the quadrature squeezing in the atomic mode for a given enhanced optomechanical coupling rate $G_1$  in two different situations: (i)  only one of the mirrors is movable (we consider the mirror 2 to be fixed, i.e., $G_2=0$) and (ii) both mirrors move. The quadrature squeezing, which results from the transfer of state from the mechanical Bogoliubov mode to the atomic mode, can be controlled by the enhanced optomechanical coupling rate $G_1$ and the atom-field coupling $\eta_e$.

From now on, for the sake of simplicity, we assume that the two mechanical oscillators are identical, i.e., ${\omega_2=\omega_1}$, ${\lambda_2=\lambda_1}$, ${g_2=g_1}$, \begin{small}${G_2=G_1}$\end{small}, and are coupled to the thermal baths with common temperature $T_2=T_1=T$. Also, both mechanical oscillators as well as the atomic and optical modes are assumed to be prepared in their corresponding ground states (${\oper o\vert0_o\rangle=0,\,\,\,o=a,c,b_j}$).

For the case (i) straightforward calculations show that in the absence of the noise sources the state transfer can take place between the mechanical mode $\toper B_1$ and the atomic mode $\toper c$ after the operating time  $t_{s_{\text{(i)}}}^{{(0)}}=\frac{\pi}{\sqrt{{G'^{2}_{1}}+\eta_{e}^2}}$, and is a perfect process when $\eta_{e}=G'_{1}$ is established. In consequence 
\begin{small}\begin{subequations}\label{kkk1}\begin{align}
&\toper c(t_{s_{\text{(i)}}}^{{(0)}})=\oper{B}_1(0),\\
&\toper B_1(t_{s_{\text{(i)}}}^{{(0)}})=\oper c(0).
\end{align}\end{subequations}\end{small}In addition, for the case (ii) it can be shown that a perfect state swapping between the atomic mode and the collective mechanical mode ${\frac{\toper{B}_1-\toper{B}_2}{\sqrt{2}}}$ can occur after the operating time  $t_{s_{\text{(ii)}}}^{{(0)}}=\frac{\pi}{\sqrt{2{G'^2_{1}}+\eta_{e}^2}}$ and for $\eta_{e}=\sqrt{2}G'_{1}$. In this case, Eqs.~(\ref{c1})-(\ref{B2}) can be rewritten as
\begin{small}\begin{subequations}\label{kkk2}\begin{align}
&\toper c(t_{s_{\text{(ii)}}}^{{(0)}})=\frac{\oper{B}_1(0)-\oper{B}_2(0)}{\sqrt{2}},\\
&\frac{\toper B_1(t_{s_{\text{(ii)}}}^{{(0)}})-\toper B_2(t_{s_{\text{(ii)}}}^{{(0)}})}{\sqrt{2}}=\oper c(0).
\end{align}\end{subequations}\end{small}The atomic quadrature squeezing is the result of the state transfer described by \cref{kkk1,kkk2}. To  justify this claim, we consider the variances of the atomic quadratures for the cases (i) and (ii) that can be obtained from Eqs.~(\ref{kkk1}) and (\ref{kkk2}) as $\text{var}(\oper x_c)=\frac{e^{2r_{1}}}{2}$ and $\text{var}(\oper y_c)=\frac{e^{-2r_{1}}}{2}$, which clearly indicate controllability of the quadrature variances through the mechanical anharmonicity.  Note that the squeezing parameter $r_1$ takes different values for the two cases (i) and (ii).

To quantify the quadrature squeezing, we consider the degree of  squeezing for the quadrature $\oper y_c$ which in the dB (decibel) unit is defined by~\cite{agar2010} ${D_{y_c}=-10 \log_{10}\frac{\text{var}(\oper y_c(t_s))}{\text{var}(\oper y_c(0))}}$ where according to the chosen initial condition for the atomic mode, we have $\text{var}(\oper y_c(0))=(\text{var}(\oper y_c))_{\text{vac}}=1/2$. Based on this definition, the quadrature $\oper y_c$ is said to be squeezed whenever $D_{y_c}>0$.
\begin{figure}[h]\centering
\includegraphics[scale=0.5]{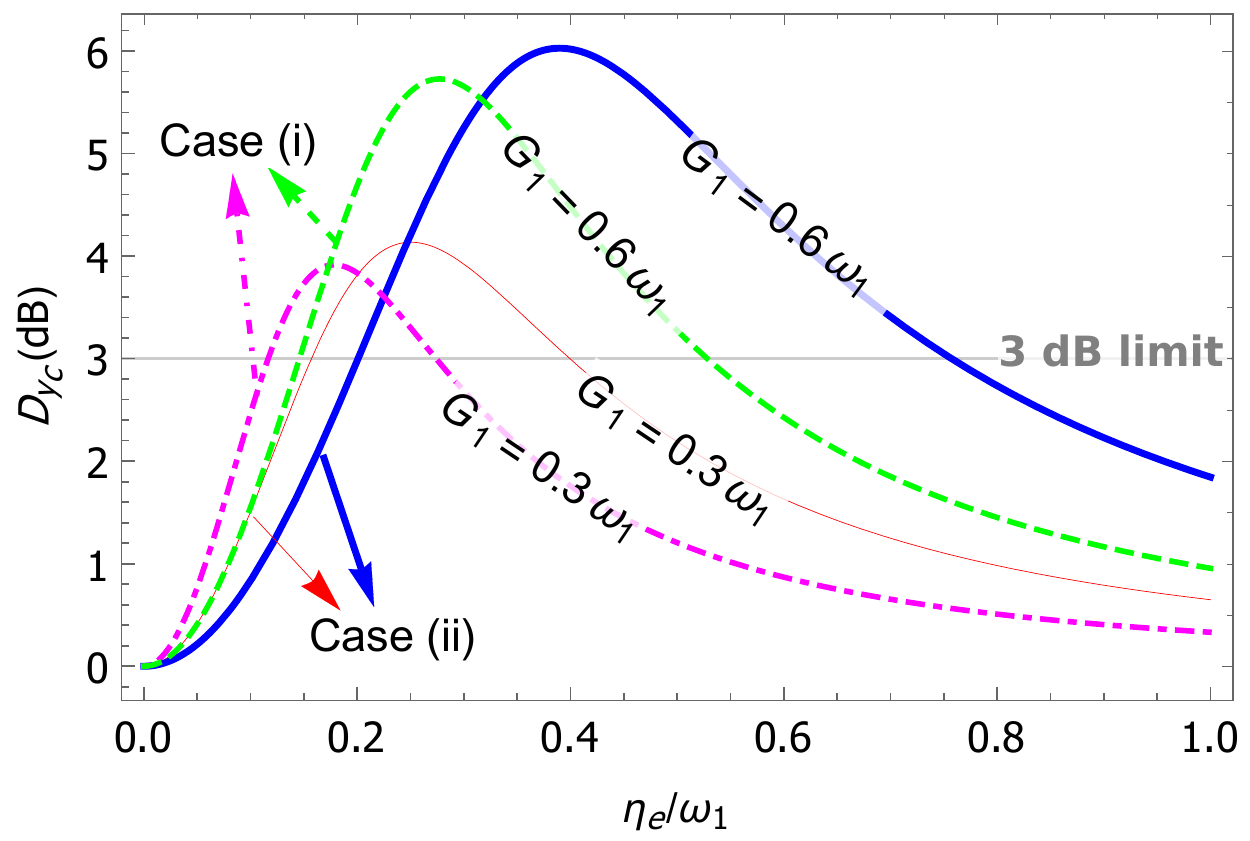}
\caption{\label{fig3} The degree of squeezing of the atomic quadrature $\oper y_c$ in dB unit versus the normalized atom-field coupling $\eta_e/\omega_1$ for two different values of the enhanced optomechanical couplig rate $G_1$ when $T=25\text{mK}$, $\lambda_1=10^{-4}\omega_1$,  $g_1=6.07\times10^{-4}\omega_1$, and $\kappa=0.01\omega_1$. For the purpose of comparison, the results of an optomechanical system with a single Duffing-like moving mirror (case (i)) and those of an optomechanical system with two Duffing-like movable mirrors (case (ii)) are plotted alongside each other. The 3 dB limit of squeezing which corresponds to 50\% noise reduction below the zero-point level has also been shown. Other parameters are the same as those in \cref{fig2}.}
\end{figure}Figure \ref{fig3} illustrates the variation of the degree of the squeezing $D_{y_c}$ with respect to $\eta_e$ for two different values of $G_1$. This figure clearly shows that, in both cases (i) and (ii), with increasing $G_1$ the generated squeezing becomes stronger, and also the quadrature squeezing beyond the standard 3 dB limit can be achieved in a wide range of values of $\eta_e$.  
It should be noted that for the cases (i) and (ii) the maximum value of the degree of the quadrature squeezing occurs, respectively, when  $\eta_e=G'_{1}$  and $\eta_e=\sqrt{2}G'_{1}$  are established. Furtheremore, comparing cases (i) and (ii) for a given value of $G_1$ indicates that the generated quadrature squeezing in case (ii) is not only stronger but also occurs in a wider range of values of  $\eta_e$.

\begin{figure}[ht]
\centering\includegraphics[scale=0.5]{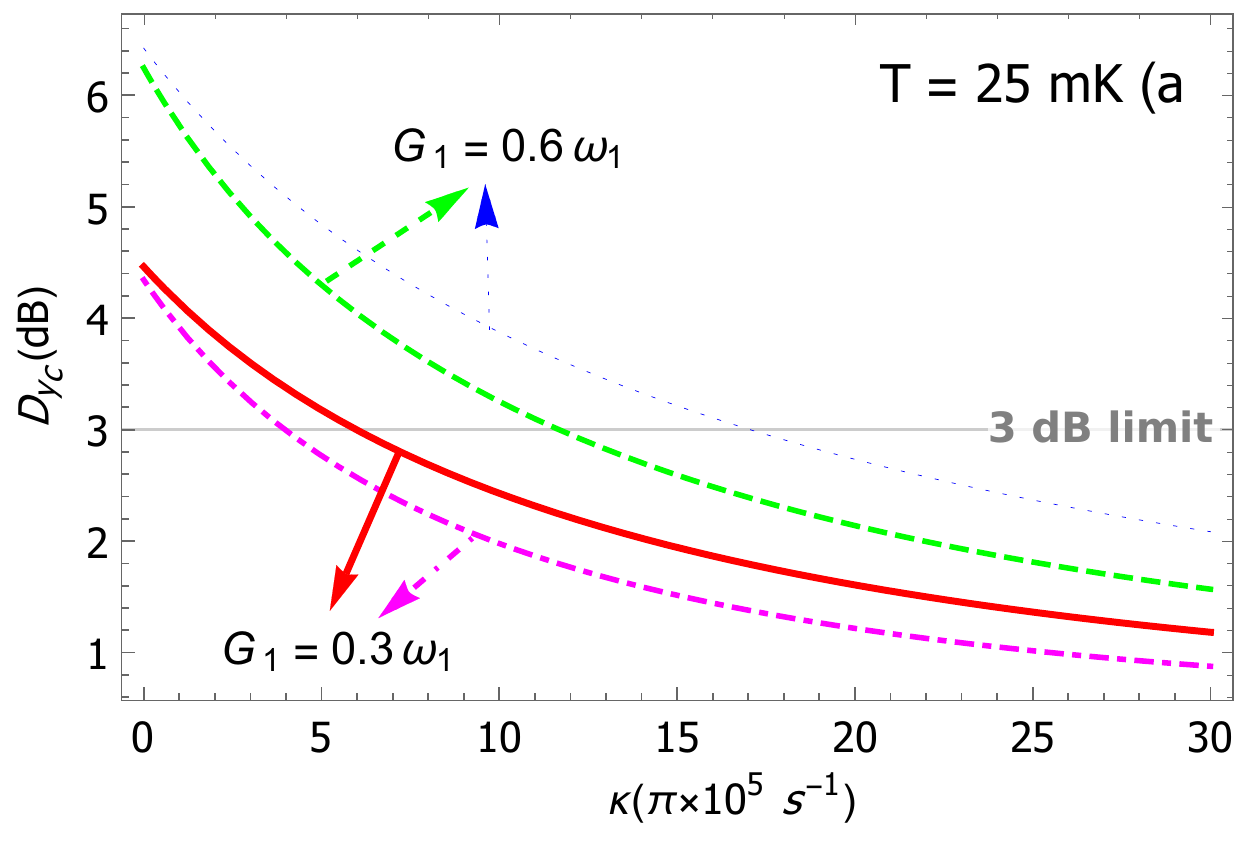}\par\includegraphics[scale=0.5]{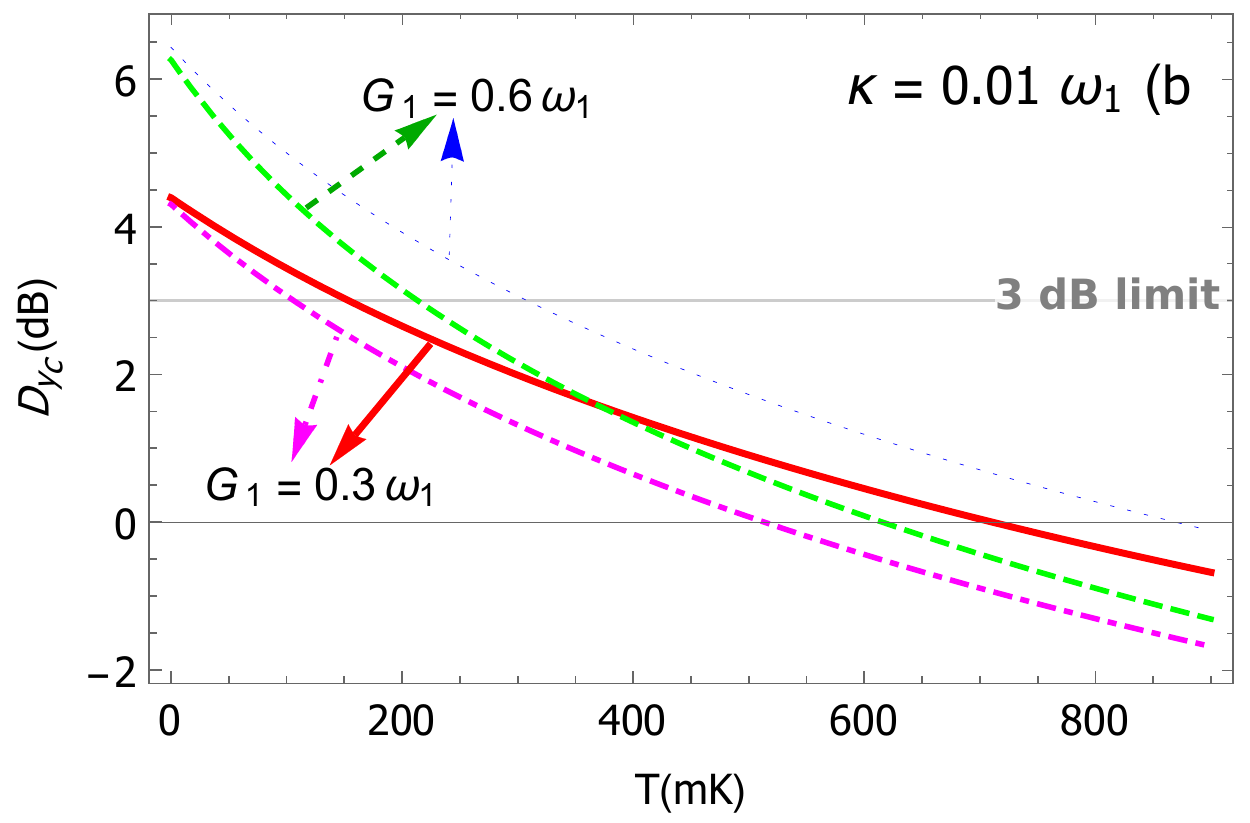}
\caption{\label{fig4} Maximum available degree of atomic quadrature squeezing shown in \cref{fig3} versus (a) the cavity damping rate $\kappa$ for $T=25\text{mK}$, and (b) the temperature of the thermal baths $T$ for $\kappa=0.01\omega_1$. The green dashed and magenta dot-dashed lines show the degree of squeezing in the optomechanical system when only one of the mirrors is movable, while the blue dotted and  red solid lines present  the same quantity in the optomechanical system with two Duffing-like movable mirrors. Other parameters are the same as those used in \cref{fig3}. }
\end{figure}

 To study the robustness of the generated squeezing against the noise sources, the behavior of the peaks of the degree of the quadrature squeezing shown in \cref{fig3} has been plotted versus the cavity damping rate $\kappa$ and temperature $T$, respectively, in Fig.~\ref{fig4}(a) and Fig.~\ref{fig4}(b). In these two figures, the green  dashed  and  magenta dot-dashed lines correspond to the case (i) while the blue dotted and  red  solid lines correspond to the case (ii). 
As can be seen, in both cases (i) and (ii), with increasing $G_1$ the generated squeezing becomes more robust. Although  in an optomechanical system with Duffing-like  mirrors  the thermal noise depends on the squeezing parameter $r_1$ (see definition of $\bar{n}_r$ given just after \cref{tcf3}) and increases with increasing $G_1$,  due to the dramatic reduction of the time needed for the state to be completely transferred, the effect of the thermal noise on the quadrature squeezing is reduced. Also, as already mentioned, the existence of the Duffing anharmonicity makes it possible to reduce the ratio $\frac{\kappa}{G'_1}$  by increasing $G'_1$ without breaking down the validity of the RWA. In addition, comparing the results of the cases (i) and (ii) shows that, for a given optomechanical coupling rate $G_1$, the degree of generated quadrature squeezing in the optomechanical system with two Duffing-like movable mirrors changes slower with $\kappa$ and $T$ than that corresponding to the system with a single Duffing-like movable mirror, which is due to the shorter transfer time in the former case.

\subsection{quantum state transfer between mechanical oscillators\label{sec32}}
Here, we assume that the state to be transferred is a squeezed state of the form $\vert \varrho,\xi\rangle=D(\varrho)S(\xi)\vert0_1\rangle$ initially mapped on the mechanical mode 1, where $D(\varrho)=\exp(\varrho \,\oper b_1^\dagger-\varrho^*\oper b_1)$ is the displacement operator with the coherent amplitude $\varrho$  and $S(\xi)=\exp(\frac{\xi}{2}(\oper b_1^2-\oper b_1^{\dagger 2}))$ is the single-mode squeezing operator with $\xi$ being the real squeezing parameter. This initial Gaussian  state is determined by the first moment 
\begin{align}
 & \langle\oper{ \vec{X}}_{\text{ini}}\rangle=(e^{r_1}\sqrt{2}\,\text{Re}(\varrho ),e^{-r_1}\sqrt{2}\,\text{Im}(\varrho))^T,\end{align} 
 together with the covariance matrix 
 \begin{align}\label{vinit} \mathbf{V}_{\text{ini}}=\frac{1}{2}\begin{pmatrix}
{e^{2r_1-2\xi}}&0\cr
0&{e^{-2r_1+2\xi}}
\end{pmatrix}.
\end{align}
In addition, we assume that the transfer destination is the mechanical mode 2 which is prepared in its ground state $\vert0_2\rangle$ ($\oper b_2\vert0_2\rangle=0$).  The atomic and cavity modes are also considered to be initially prepared in their corresponding ground states. The final Gaussian state is determined by 
the first moment (see \cref{key1})
 \begin{align}
& \langle\oper{ \vec{X}}_{\text{fin}}\rangle =M_{5,7}(t_s) \langle\oper{ \vec{X}}_{\text{ini}}\rangle,
\end{align}
together with the covariance matrix 
 \begin{align}\label{vfin}
  &\mathbf{V}_{\text{fin}}(t_s)=\begin{pmatrix}
\text{var}(\oper Q_2)&\text{cov}(\oper Q_2,\oper P_2)\cr
\text{cov}(\oper Q_2,\oper P_2)&\text{var}(\oper P_2)
\end{pmatrix},
\end{align}
derived from \cref{k4} with $\mathbf{N}\approx\mathbf{N}_0$ and the definitions \begin{small}$\text{cov}(o_1,o_2)=1/2\langle o_1o_2+o_2o_1\rangle-\langle o_1\rangle\langle o_2\rangle$\end{small} and \begin{small}$\text{var}(o_1)=\text{cov}(o_1,o_1)$\end{small}.

Furthermore, in order to investigate the performance of the state transfer, we use the Ulhmann fidelity \cite{uh1976} which for the transfer of a single-mode Gaussian state can be written as \cite{T22}
\begin{align}
\mathit{F}=\frac{1}{1+\bar{n}_h}e^{-\frac{\lambda_{h}^2}{1+\bar{n}_h}},
\end{align}
with the \textit{heating} parameter $\bar{n}_h$ and \textit{amplitude-decay}  parameter $\lambda_{h}$ defined by 
\begin{small}
\begin{subequations}
\begin{align}
\bar{n}_h&=\sqrt{\det (\mathbf{V}_{\text{ini}}+ \mathbf{V}_{\text{fin}})}-1,\label{heatp}\\
\lambda_{h}^2&=(\langle \oper{\vec{X}}_{\text{fin}}\rangle-\langle \oper{\vec{X}}_{\text{ini}}\rangle).\frac{\sqrt{\det (\mathbf{V}_{\text{ini}}+ \mathbf{V}_{\text{fin}})}}{\mathbf{V}_{\text{ini}}+\mathbf{V}_{\text{fin}}}.(\langle \oper{\vec{X}}_{\text{fin}}\rangle-\langle \oper{\vec{X}}_{\text{ini}}\rangle).
\end{align}
\end{subequations}
\end{small}In order to achieve a perfect state transfer both parameters  $\bar{n}_h$ and  $\lambda_{h}$ must be reduced as much as possible.

\begin{figure}[h]
\centering\includegraphics[scale=0.4]{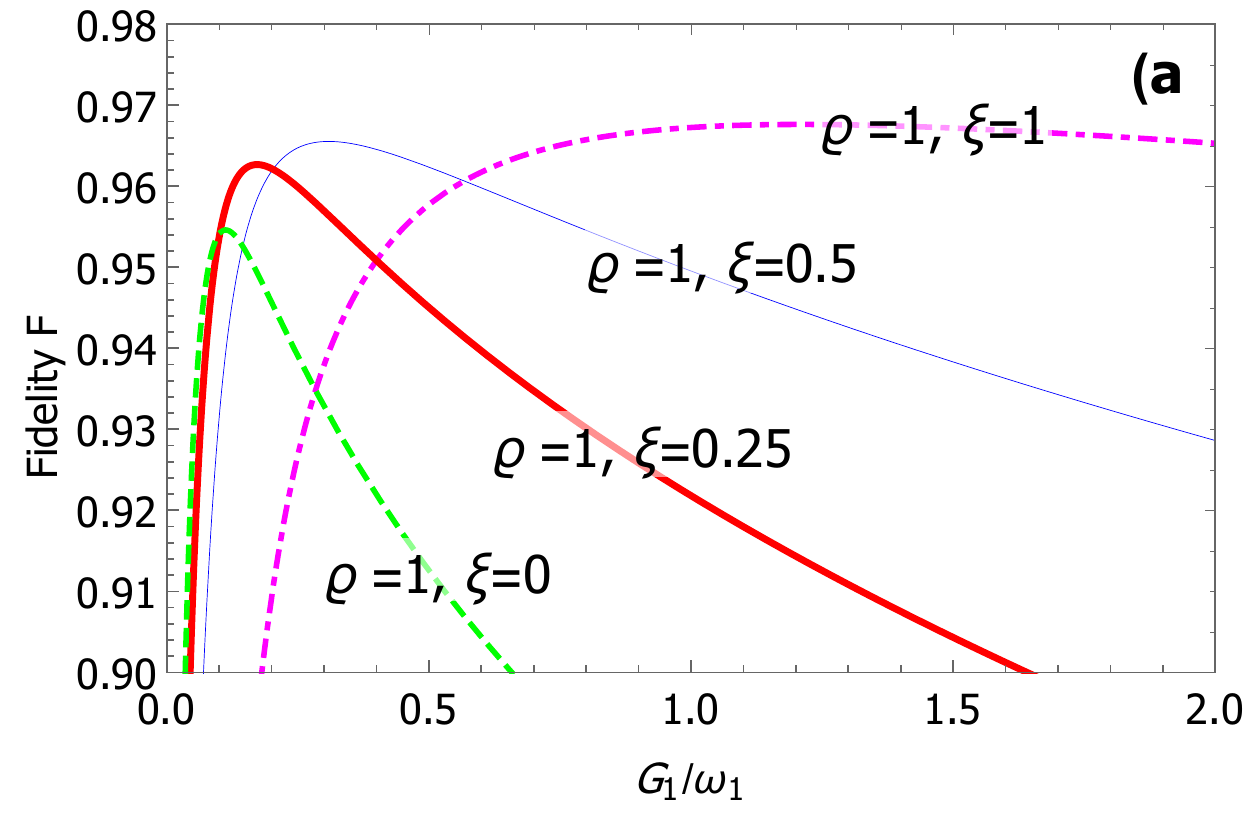}\par\includegraphics[scale=0.4]{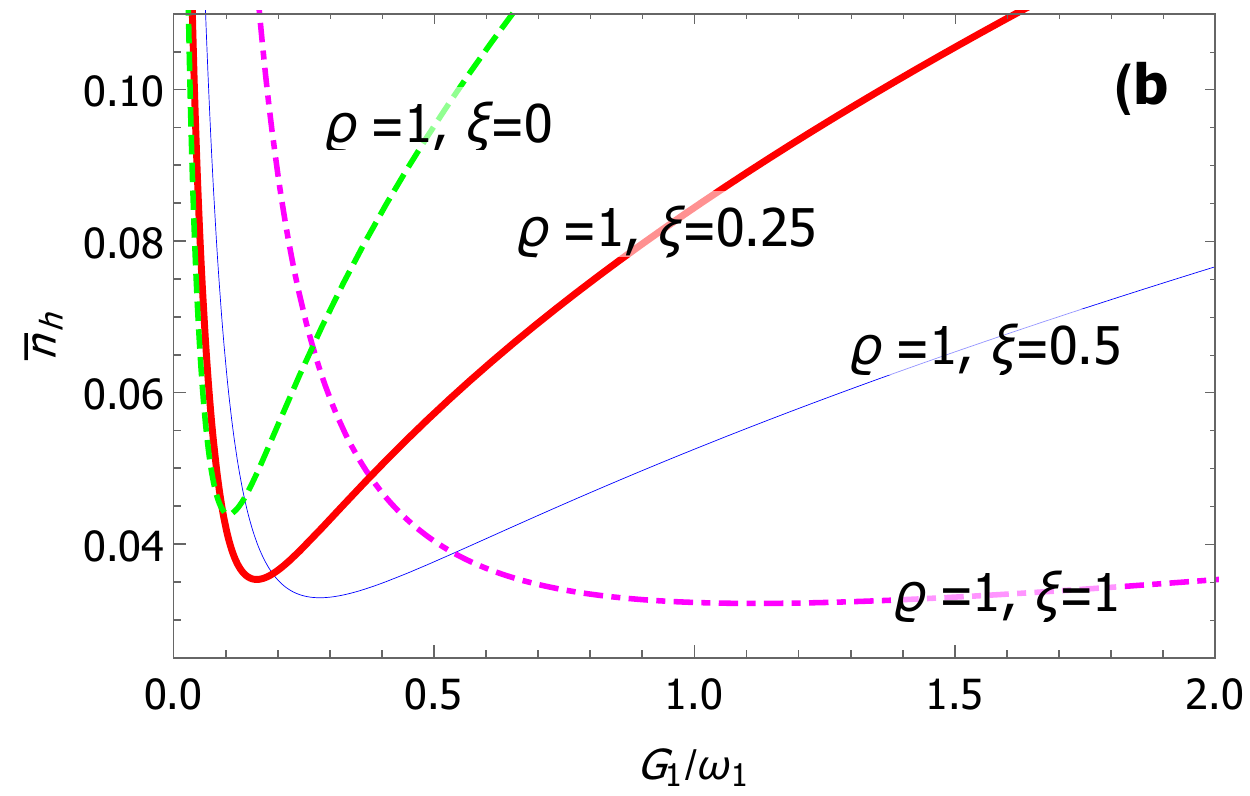}\par\includegraphics[scale=0.4]{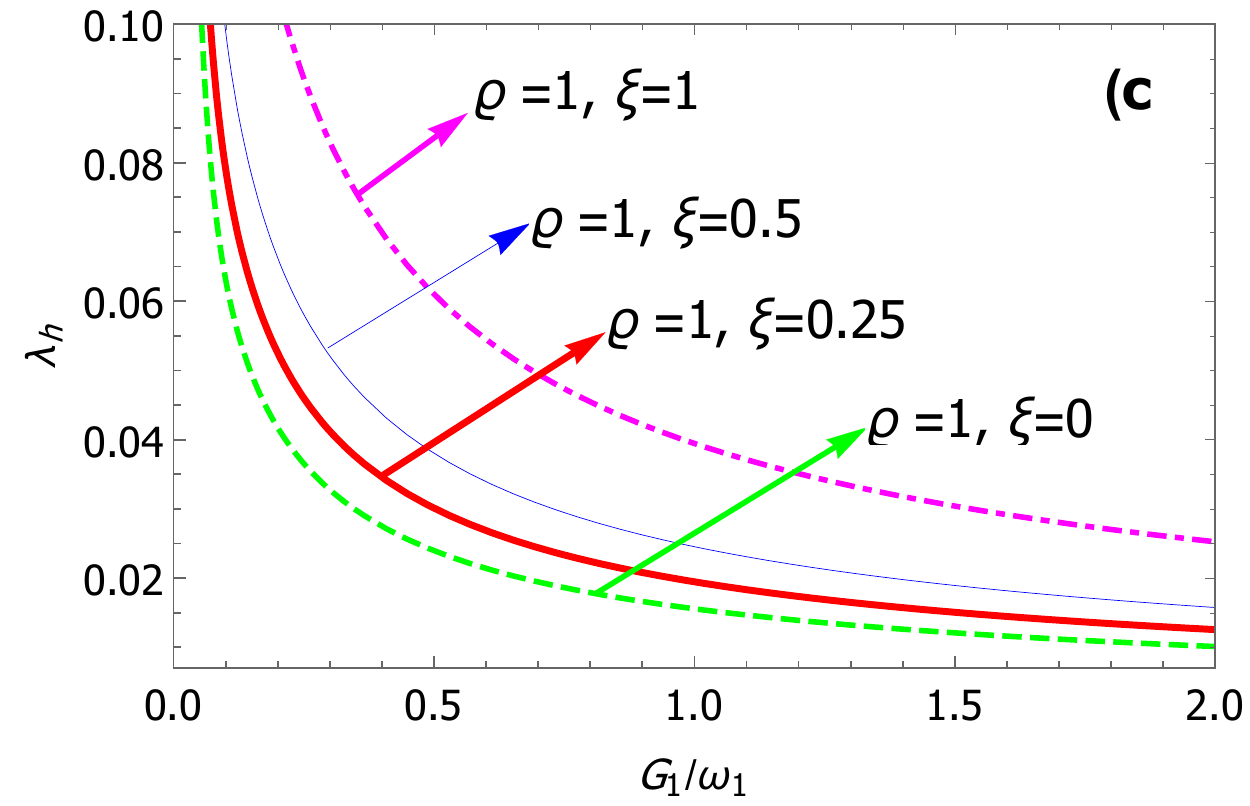}
\caption{\label{fig5} (a) The transfer fidelity $F$, (b) the heating parameter $\bar{n}_h$, and (c)  the amplitude-decay parameter $\lambda
_h$ versus the normalized enhanced optomechanical coupling rate $G_1/\omega_1$ for different initial states $|\varrho,\xi\rangle$. The parameters used in this figure are:  $\kappa=0.01\omega_1$, $\eta_e=0.005\omega_1$, $T=25\text{mK}$, $\lambda_1=10^{-4}\omega_1$, and $g_1=6.07\times10^{-4}\omega_1$. Other parameters are the same as those in \cref{fig4}. }
\end{figure}

 The performance of the state transferring (fidelity)  as a function of  the normalized enhanced optomechanical coupling rate $G_1/\omega_1$ for different values of $\varrho$ and $\xi$ has been plotted in \cref{fig5}(a). 
In addition, the variations of $\bar{n}_h$ and $\lambda_h$ with $G_1/\omega_1$ have been plotted, respectively, in \cref{fig5}(b) and \cref{fig5}(c). It is well seen from Fig.~\ref{fig5}(a) that there is an optimal value of $G_{1}$, denoted by $G_{1,\text{opt}}$, for which the fidelity is maximal. Therefore, although the presence of the Duffing anharmonicity extends the range of validity of the RWA, this does not mean that  if we increase $G_1$ to any desired value, the maximum available fidelity  will also increase. The maximum available  $F$ as well as the width of the peak of the fidelity  depend on the magnitude of the squeezing parameter of the initial state written on the mechanical mode $\toper B_1$. Fidelity reaches its  maximum value when $\bar{n}_h$ and $\lambda_h$ are reduced as much as possible. Numerical calculation shows that the minimum of $\bar{n}_h$ located at the solution of $r_1\big|_{G_1=G_{1,\text{opt}}}\approx\xi+\frac{\pi\,\kappa}{2\mathcal{G}}\big|_{G_1=G_{1,\text{opt}}}$, approximately determines the maximum of $F$. So, in addition to $\xi$, the value of $G_{1,\text{opt}}$ depends also on the cavity damping rate, atom-field coupling strength, and the Duffing parameter $\lambda_1$. Since $r_1$ is an ascending function of $G_1$, with increasing $\xi$ the point of intersection between $r_1$ and $\xi+\frac{\pi\,\kappa}{2\mathcal{G}}$ ($G_{1,\text{opt}}$) is shifted to larger values of $G_1$. By increasing $G_{1,\text{opt}}$ the time needed to transfer the state and the contributions of the noises to the degradation of the state are noticeably reduced.  Consequently, by tuning $G_1=G_{1,\text{opt}}$ a given quantum state can be transferred with the highest possible fidelity. 
\begin{figure}
\centering\includegraphics[scale=0.4]{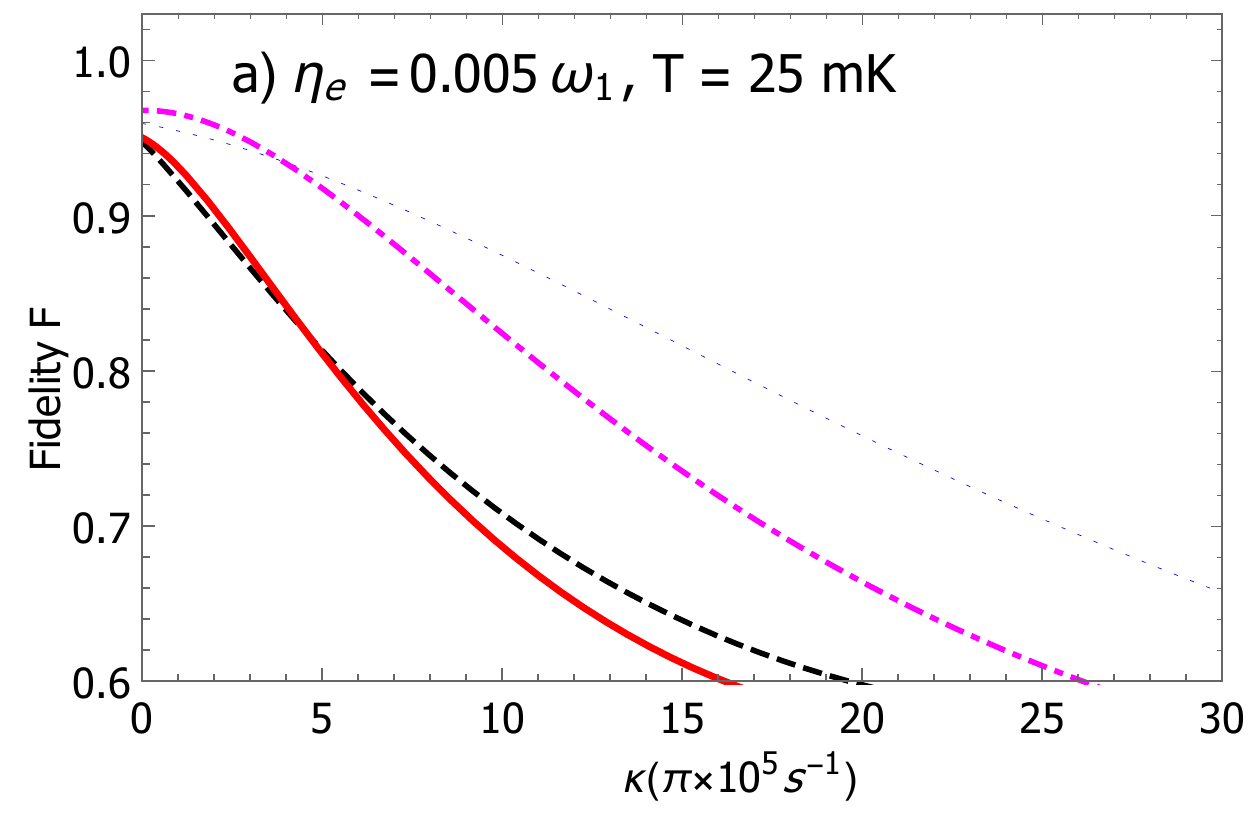}\par\includegraphics[scale=0.4]{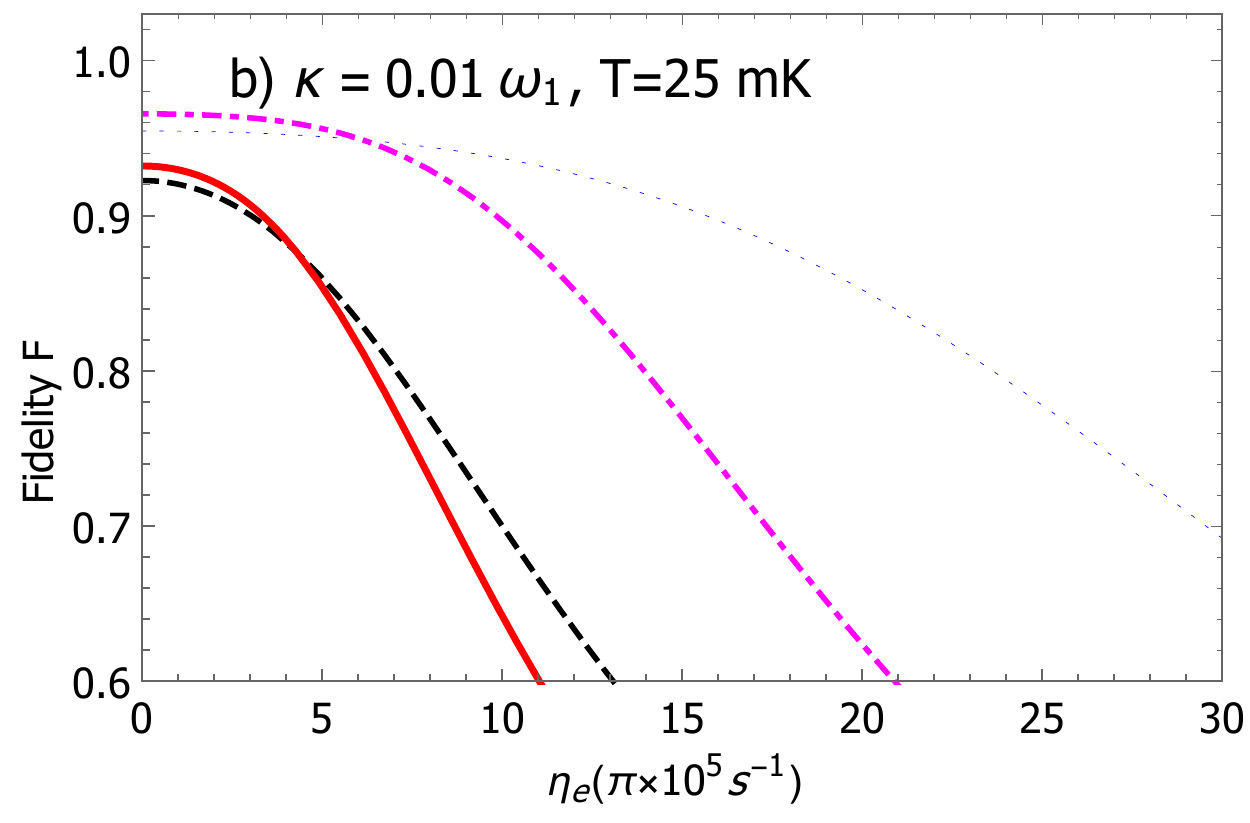}\par\includegraphics[scale=0.4]{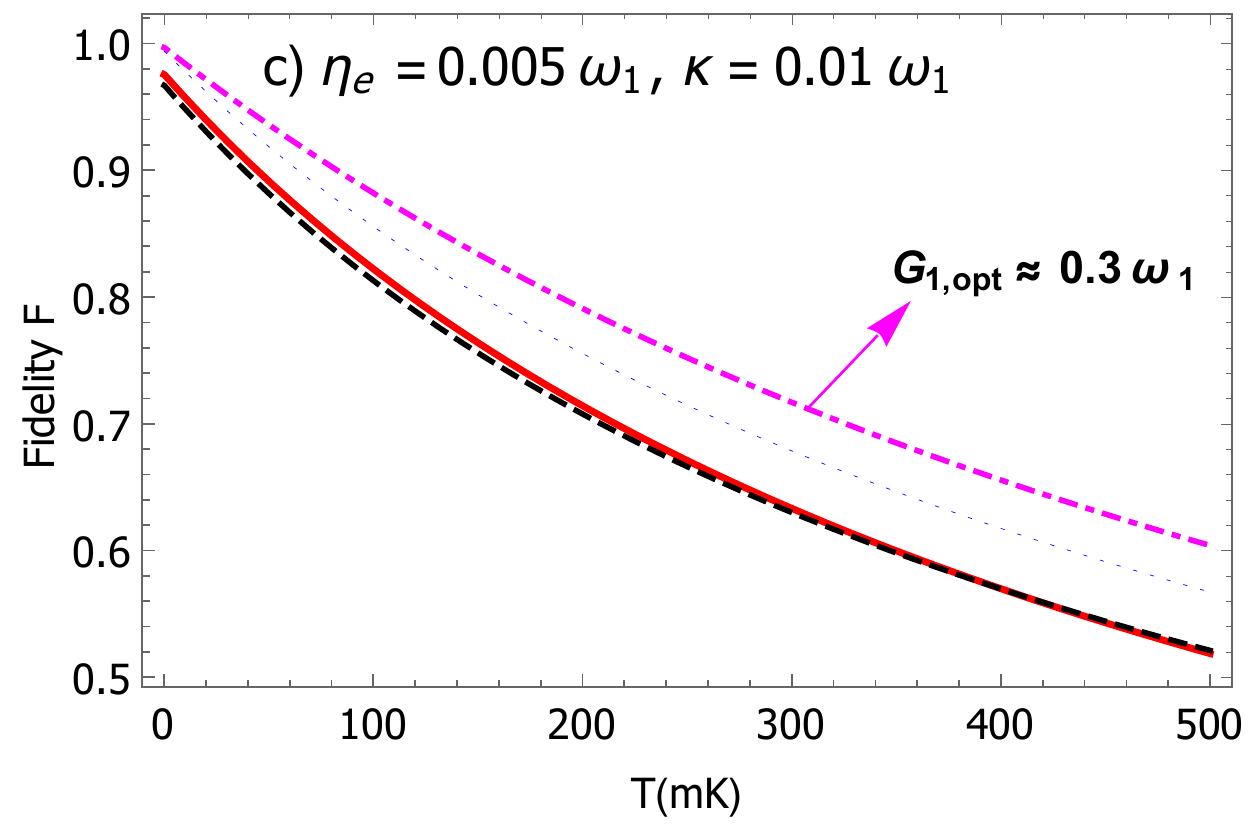}
\caption{\label{fig6} The fidelity $F$ for the transfer of the squeezed state $|\varrho=1,\xi=0.5\rangle$  versus the (a) cavity damping rate $\kappa$, (b) enhanced atom-field coupling $\eta_e$, and (c) common temperature of the mechanical baths $T$ for different values of $G_1$: $G_1/\omega_1=0.1$ (red solid line), $G_1/\omega_1=0.3$ (magenta dot-dashed line ), $G_1/\omega_1=0.8$ (blue dotted line). In each panel, black dashed line shows the transfer fidelity in the absence of Duffing anharmonicity ($\lambda_1=\lambda_2=0$) for  $G_1/\omega_1=0.1$.  Other parameters are the same as those in \cref{fig5}. }
\end{figure}
Another point to be noted here is that for $G_1\ll\omega_1$, quantum states with larger squeezing parameters are transferred with lower fidelities. This illustrates well that the system under investigation  is a good candidate for the transfer of strongly squeezed quantum states.

To investigate the effects of the sources that disrupt the fidelity ($\kappa$,$\eta_e$,$T$), in Figs.~\ref{fig6}(a)-\ref{fig6}(c) we have plotted the variation of the fidelity with respect to  the cavity damping rate,  the atom-field coupling, and the temperature of the baths  for the initial squeezed state $|\varrho=1,\xi=0.5\rangle$. For comparison, the fidelity of the state transferred from the mechanical oscillator 1 to the mechanical oscillator 2 in the absence of the mechanical anharmonicity ($\lambda_2=\lambda_1=0$)  for $G_1=0.1\omega_1$ (where the RWA is valid) has also been shown  in each panel (black dashed line).

As previously mentioned, if $\eta_e$ is weaker than $\sqrt{2}G'_1$, then the quantum state will be exchanged between the two mechanical oscillators with a higher fidelity. In addition, by decreasing the ratio $\kappa/G'_1$  the time needed to transfer the quantum state and, consequently, the destructive effect of the cavity damping will be reduced. As a result of these facts, for stronger optomechanical coupling rate $G_1$ the fidelity $F$ reduces more slowly with increasing  $\kappa$ (\cref{fig6}(a)) and $\eta_e$ (\cref{fig6}(b)). It should be noted that in the panel (c) the magenta dot-dashed  line shows the variation of the fidelity  for $G_1=G_{1,\text{opt}}\approx0.3\omega_1$. As can be seen from the figure, the fidelity for $G_1=G_{1,\text{opt}}$ exhibits the highest robustness to the increase in the temperature. In the system under consideration, the thermal phonon number $\bar{n}_{2,r}=\bar{n}_{1,r}=\bar{n}_{r}$ is an ascending function of $G_1$ and $T$. For this reason, when $G_1>G_{1,\text{opt}}$  the fidelity decreases rapidly with increasing $T$.

\section{conclusion\label{conc}}
In conclusion, we have studied theoretically the quantum state transfer in a laser driven hybrid optomechanical cavity with two Duffing-like movable end mirrors containing an atomic ensemble for the case before the cavity field reaches the steady state.  We have explored the squeezing transfer from the collective mechanical Bogoliubov mode to the atomic ensemble as well as the transfer process of a pre-written squeezed state from one mechanical mode to another. We have realized that the state transfer from the Bogoliubov modes of the anharmonic oscillators to the atomic mode may lead to the controllable atomic quadrature squeezing beyond the standard quantum limit of 3 dB. In particular, the results reveal that the robustness of the generated atomic quadrature squeezing against the noise sources can be enhanced by the Duffing anharmonicity. Besides, we have shown that in the presence of the Duffing anharmonicity  it is possible to transfer strongly squeezed states between the two mechanical oscillator in a short operating time and with a high fidelity.  This suggests that the Duffing anharmonic mechanical oscillators are more suitable candidates as the nodes in quantum networks desingned for transferring strongly squeezed states.

\section*{Appendix: The explicit forms of the matrix $\mathbf{M}$, vector of noise operators  $\mathbf{\Gamma}$, and diffusion matrix $\mathbf{N}$}
\setcounter{equation}{0}
\renewcommand{\theequation}{A\arabic{equation}}
The equations of motion (\ref{eq1})-(\ref{eq111}) lead to the following equations of motion for the vector of quadrature fluctuation operators $\oper{\vec{ U}}$ in the compact matrix form
\begin{align}\label{key2}
&\delta\dot{\hat{\vec{ U}}}(t)=\mathbf{A}.\oper{\vec{ U}}(t)+\hat{\vec{ n}}(t),
\end{align}
where $\hat{\vec{ n}}=(\sqrt{2\gamma}\,\hat{ x}_{in,c},\sqrt{2\gamma}\,\hat{y}_{in,c},\sqrt{2\kappa}\,\hat{ x}_{in,a},\sqrt{2\kappa}\,\hat{y}_{in,a},$\\$\sqrt{2\gamma}\,\hat{Q}_{in,1},\sqrt{2\gamma}\,\hat{P}_{in,1},\sqrt{2\gamma}\,\hat{Q}_{in,2},\sqrt{2\gamma}\,\hat{P}_{in,2})^T$ is the vector of noises in which $\hat{ x}_{in,c}=\frac{\tilde{c}_{in}+\tilde{c}_{in}^\dagger}{\sqrt{2}}$ and $\hat {y}_{in,c}=\frac{\tilde{c}_{in}-\tilde{c}_{in}^\dagger}{\sqrt{2}i}$ represent the input noise quadratures of the atomic ensemble, $\hat{ x}_{in,a}=\frac{\tilde{a}_{in}+\tilde{a}_{in}^\dagger}{\sqrt{2}}$ and $\hat {y}_{in,a}=\frac{\tilde{a}_{in}-\tilde{a}_{in}^\dagger}{\sqrt{2}i}$ denote the input noise quadratures of the optical field, and  $\hat{ Q}_{in,j}=\frac{\tilde{B}_{in,j}+\tilde{B}_{in,j}^\dagger}{\sqrt{2}}$ and $\hat{P}_{in,j}=\frac{\tilde{B}_{in,j}-\tilde{B}_{in,j}^\dagger}{\sqrt{2}i}$ ($j=1,2$) refer to the input noise quadratures of the mechanical mode $j$. Also,  the \textit{drift} matrix $\mathbf{A}$ is given by
 \begin{footnotesize}
 \begin{align} \mathbf{A}=\begin{pmatrix}
-\gamma&0&0&\eta_e&0&0&0&0\cr
0&-\gamma&-\eta_e&0&0&0&0&0\cr
0&\eta_e&-\kappa &0&0&-G_1'&0&G_2'\cr
-\eta_e&0&0&-\kappa&G_1'&0&-G_2'&0\cr
0&0&0&-G_1'&-\gamma&0&0&0\cr
0&0&G_1'&0&0&-\gamma&0&0\cr
0&0&0&G_2'&0&0&-\gamma&0\cr
0&0&-G_2'&0&0&0&0&-\gamma
\end{pmatrix}.
\end{align}\end{footnotesize}
By applying the Laplace transform defined by
\begin{equation}\label{LT}
\bar{O}(s)=\int_0^\infty\,dt\,e^{-s\,t}\hat{O}(t).
\end{equation}
Eq.~(\ref{key2}) can be written as
\begin{footnotesize}\begin{align}
\delta\bar{\vec{ U}}(s)&=\mathcal{M}(s).\delta{\vec{ U}}(t_0=0)+ \mathcal{M}(s).\bar{\vec{ n}}(s),
\end{align}\end{footnotesize}
with
\begin{footnotesize}
\begin{equation}
\mathcal{M}(s)=(s \,\,\mathbf{I}_{8\times 8}-\mathbf{A})^{-1} .
\end{equation}\end{footnotesize}
Performing  the inverse Laplace transform defined by 
\begin{equation}\label{ILT}
\hat{O}(t)=\frac{1}{2\pi\,i}\lim_{\tau\rightarrow\infty}\int_{s_r-i\,\tau}^{s_r+i\,\tau}ds\, e^{s\, t}\bar{O}(s),
\end{equation}

and reusing of \cref{LT} lead to
\begin{align}
\oper{\vec{ U}}(t)&=\mathbf{M}(t).\oper{\vec{ U}}(t_0=0)+\hat{\vec{\Gamma}}(t),\end{align}
where 
\begin{equation}
\hat{\vec{\Gamma}}(t)=\int_{t_0=0}^t d\tau\, \mathbf{M}(t-\tau).\hat{\vec{ n}}(\tau),
\end{equation}
and 
\begin{footnotesize}
\begin{align}
\mathbf{M}(t)&=\frac{1}{2\pi\,i}\lim_{\tau\rightarrow\infty}\int_{s_r-i\,\tau}^{s_r+i\,\tau}ds\, e^{s\, t}\mathcal{M}(s)\nonumber\\
\equiv &\begin{pmatrix}
M_{1,1}&0&0&M_{1,4}&M_{1,5}&0&M_{1,7}&0\cr
0&M_{1,1}&-M_{1,4}&0&0&M_{1,5}&0&M_{1,7}\cr
0&M_{1,4}&M_{3,3}&0&0&M_{3,6}&0&M_{3,8}\cr
-M_{1,4}&0&0&M_{3,3}&-M_{3,6}&0&-M_{3,8}&0\cr
M_{1,5}&0&0&M_{3,6}&M_{5,5}&0&M_{5,7}&0\cr
0&M_{1,5}&-M_{3,6}&0&0&M_{5,5}&0&M_{5,7}\cr
M_{1,7}&0&0&M_{3,8}&M_{5,7}&0&M_{7,7}&0\cr
0&M_{1,7}&-M_{3,8}&0&0&M_{5,7}&0&M_{7,7}
\end{pmatrix},
\end{align}\end{footnotesize}with $s_r$ being the real part of variable $s$. By straightforward calculation, the entries of the matrix $\mathbf{M}$ ($M_{i,j}$) can be obtained as
\begin{footnotesize}
\begin{subequations}
\begin{align}
& \hspace*{-0.5cm}M_{1,1}=\frac{e^{-\chi_1t}}{G'^{2}_1+G'^{2}_2+\eta_e^2}\Big(e^{\chi_2t}(G'^{2}_1+G'^{2}_2)+\eta_e^2\cos(\mathcal{G}t)+\frac{\eta_e^2\chi_2}{\mathcal{G}}\sin(\mathcal{G}t)\Big),\\
& \hspace*{-0.5cm}M_{1,4}=e^{-\chi_1t}\frac{\eta_e}{\mathcal{G}}\sin(\mathcal{G}t),\\
&\hspace*{-0.5cm} M_{1,5}=-\frac{G_1'\eta_e e^{-\chi_1t}\Big(\cos(\mathcal{G}t)-e^{\chi_2t}+\frac{\chi_2}{\mathcal{G}}\sin(\mathcal{G}t)\Big)}{G'^{2}_1+G'^{2}_2+\eta_e^2},\\
&\hspace*{-0.5cm} M_{1,7}=\frac{G_2'\eta_e e^{-\chi_1t}\Big(\cos(\mathcal{G}t)-e^{\chi_2t}+\frac{\chi_2}{\mathcal{G}}\sin(\mathcal{G}t)\Big)}{G'^{2}_1+G'^{2}_2+\eta_e^2},\\
&\hspace*{-0.5cm} M_{3,3}=e^{-\chi_1t}(\cos(\mathcal{G}t)-\frac{\chi_2}{\mathcal{G}}\sin(\mathcal{G}t)),\\
& \hspace*{-0.5cm}M_{3,6}=-e^{-\chi_1t}\frac{G_1'}{\mathcal{G}}\sin(\mathcal{G}t),\\
& \hspace*{-0.5cm}M_{3,8}=e^{-\chi_1t}\frac{G_2'}{\mathcal{G}}\sin(\mathcal{G}t),\\
& \hspace*{-0.5cm}M_{5,5}=\frac{e^{-\chi_1t}}{G'^{2}_1+G'^{2}_2+\eta_e^2}\Big(e^{\chi_2t}(\eta_e^2+G'^{2}_2)+G'^{2}_1\cos(\mathcal{G}t)+\frac{G'^{2}_1\chi_2}{\mathcal{G}}\sin(\mathcal{G}t)\Big),\\
&\hspace{-0.5cm}M_{5,7}=-\frac{G_1'G_2' e^{-\chi_1t}\Big(\cos(\mathcal{G}t)-e^{\chi_2t}+\frac{\chi_2}{\mathcal{G}}\sin(\mathcal{G}t)\Big)}{G'^{2}_1+G'^{2}_2+\eta_e^2},\\
& \hspace*{-0.5cm}M_{7,7}=\frac{e^{-\chi_1t}}{G'^{2}_1+G'^{2}_2+\eta_e^2}\Big(e^{\chi_2t}(\eta_e^2+G'^{2}_1)+G'^{2}_2\cos(\mathcal{G}t)+\frac{G'^{2}_2\chi_2}{\mathcal{G}}\sin(\mathcal{G}t)\Big),
\end{align}\end{subequations}\end{footnotesize}where $\chi_1=\frac{\kappa+\gamma}{2}$, $\chi_2=\frac{\kappa-\gamma}{2}$, and $\mathcal{G}=\sqrt{G'^{2}_1+G'^{2}_2+\eta_e^2-\chi_2^2}$.

Finally, straightforward calculation shows that the diffusion matrix $\mathbf{N}$, introduced in the text,  can be written as the sum of two distinc parts: the time-independent diagonal matrix $\mathbf{N}_0$ given by
\begin{footnotesize}
\begin{align}
\mathbf{N}_0&=\text{Diag}\{\gamma,\gamma,\kappa,\kappa,\gamma(1+2\bar{n}_{r,1}),\nonumber\\&\hspace{1.42cm}\gamma(1+2\bar{n}_{r,1}),\gamma(1+2\bar{n}_{r,2}),\gamma(1+2\bar{n}_{r,2})\},
\end{align}\end{footnotesize}and the time-dependent part $\mathbf{N}_t$ given by
\begin{footnotesize}
\begin{align}
\mathbf{N}_t&=\begin{pmatrix}
\mathbf{0}_{4\times 4}&\mathbf{0}_{4\times 4}\cr
\mathbf{0}_{4\times 4}&\mathcal{N}_{th}
\end{pmatrix},
\end{align}\end{footnotesize}with
\begin{footnotesize}
\begin{align}
\mathcal{N}_{th}&=\begin{pmatrix}
\mathcal{N}_1\cos(2\Omega_1t)&\mathcal{N}_1\sin(2\Omega_1t)&0&0\cr
\mathcal{N}_1\sin(2\Omega_1t)&-\mathcal{N}_1\cos(2\Omega_1t)&0&0\cr
0&0&\mathcal{N}_2\cos(2\Omega_1t)&\mathcal{N}_2\sin(2\Omega_1t)\cr
0&0&\mathcal{N}_2\sin(2\Omega_1t)&-\mathcal{N}_2\cos(2\Omega_1t)
\end{pmatrix},
\end{align}\end{footnotesize}and $\mathcal{N}_j=\gamma\sinh(2r_j)(1+2\bar{n}_{th,j})$.

\end{document}